# Mobility and Social Efficiency

Ryan S. Kostiuk
University of Calgary

# I.  Introduction

Mobility, as defined by the World Health Organization's International Classification of Function and Disability (ICF), is "moving or manipulating objects, by walking, running or climbing, and by using various forms of transportation".[1] Oxford introduces the concept of impairment as "loss or abnormality of a body structure or a physiological function".[2] Mobility Aids are defined, by Ridolfo and Ward, as "devices that allow individuals with mobility impairments to be ambulatory or move around including canes, crutches, walkers, wheelchairs, scooters, prosthetics and braces".[3] Mosby's Medical Dictionary classifies a caregiver as "One who contributes to the benefits of medical, social, economic or environmental resources to a dependent person".[4] This paper will show how people with mobility impairments use mobility aides and caregivers as inputs in a production process to attain a good called mobility. It will then discuss some of the economy-wide efficiency issues that prevail within most social programs aimed at helping people with mobility impairments attain mobility.

# II.  Mobility Endowments, Mobility Functions and Utility

The Oxford definition cited above implies that someone without a mobility impairment is not lacking any mobility, by virtue of having full bodily function. This paper describes such a person as having a complete endowment of *organic mobility*. Those lacking mobility, however, by virtue of their incipient bodily structures and functions, must top up their incomplete organic endowments with *synthetic mobility*, which means that they use caregivers and mobility aides to assist them in creating mobility.

**Deriving a Mobility Function From a Mobility Endowment**

Mobility endowments, whether organic or synthetic, can be expressed mathematically as shown in Equation 1.

(1) $\quad M = m_1 + m_2 + m_3 + \ldots + m_n = \sum m$

The *mobility endowment* $M$ is therefore a sum of $n$ terms, which describe tasks or activities called *mobility goods*. A typical bundle of mobility goods might include $m_1$ as a shower in the morning, $m_2$ as getting dressed, $m_3$ and brushing teeth, $m_4$ as making breakfast, $m_5$ as eating it, and so on.

---

[1] World Health Organization, *International Classification of Functioning, Disability and Health* (Geneva, 2001).

[2] Michael P. Barnes and Anthony W. Ward, *Oxford Handbook of Rehabilitation Medicine* (New York: Oxford University Press, 2005).

[3] Heather Ridolfo and Brian W. Ward, *Mobility Impairment and the Construction of Identity (*Boulder CO: Lynne Rienner, 2013).

[4] Jeff Keith, Patricia Novak, and Michelle Elliot, eds., *Mosby's Medical, Nursing, & Allied Health Dictionary,* (New York: Mosby, 2002).

The bundling of mobility goods into mobility endowments is not very interesting when describing someone with full bodily function. For someone lacking bodily function, each mobility good, by necessity, is a function of labor and capital. Suppose that $M$ denotes a mobility endowment for such a person.

$$(2) \quad M = \sum_{i=1}^{n}\sum_{j=1}^{n}\sum_{k=1}^{n}(m_i + m_j + m_k) \quad \text{where } m_i = L^\alpha K^\beta, \ m_j = l^\gamma, \text{ and } m_k = k^\delta$$

Note that we have divided the mobility goods into three categories, and summed them to get:

$$M = \sum_{i=1}^{n}\sum_{j=1}^{n}\sum_{k=1}^{n}(m_i + m_j + m_k) = L^\alpha K^\beta + l^\gamma + k^\delta + L^\alpha K^\beta + l^\gamma + k^\delta \ldots = aL^\alpha K^\beta + bl^\gamma + ck^\delta$$

$$(3) \quad M(L,K) = aL^\alpha K^\beta + bl^\gamma + ck^\delta$$

What is the meaning of Equations 2 and 3? This is best illustrated with an example. Suppose we have a mobility creator who suffers a stroke. We would proceed by computing the sum of mobility goods in the endowment, as is illustrated in the lines immediately following Equation 2. The day would begin with a shower. The caregiver would come to wake the mobility creator, and use a mechanical lift to transfer he or she into a commode shower chair. The caregiver would then bathe the mobility creator, and then use the mechanical lift a second time to transfer the creator back to bed to be dressed. The first two $m_i$ terms capture the getting out of bed and showering goods because this employs the simultaneous use of labor and capital (in the form of caregivers mechanical lifts and commode chairs). The third $m_i$ term denotes using the mechanical lift to put the creator back on the bed before being dressed. The first $m_j$ term denotes getting dressed, because the caregiver is not using any mechanized devices to put the creator's clothes on. The $m_k$ terms describe everything the creator does after they are in their wheelchair for the day and the caregiver leaves, like watching TV, for example.

The superscript terms $\alpha$, $\beta$, $\gamma$ and $\delta$ are called the *organic parameters.* they describe the severity of the creator's impairment. The smaller they are, the more severe the impairment is. Someone with relatively large organic parameters would require relatively small quantities of labor and capital to attain a desired mobility good, while someone with small organic parameters would require a relatively large quantity of inputs to attain the same good. More will be said about these in Section IV.

**Mobility Preference Functions**

One might wonder about the utility that mobility goods generate for a creator. A *Mobility Preference Function* is a utility function that describes utility as a function of mobility and all other goods, as defined by Equation 4.

$$(4) \quad U(M,A) = M^\phi A^{1-\phi} \qquad \text{where } M = \text{mobility and } A = \text{all other goods}$$

Mobility preference functions treat a mobility endowment as a sum of monetary expenditures. For example, one could look at the shower-dress bundle introduced above and express it in terms of a dollar amount of labor and machine hours spent attaining it. Expressing these things in terms of dollars means that it is possible to compare mobility and non-mobility expenditures by how much utility they generate. This paper will refer to a creator's endowment, or total endowment, as the summed monetary value of all mobility and non-mobility goods available to them. Equation 5 expresses this mathematically, where $E_T$ is the total endowment, $E_M$ is the mobility endowment and $E_A$ is the non-mobility endowment (or the all other goods endowment).

(5) $\quad E_T = E_M + E_A$

Equation 5 gives rise to the concept of a *composite bundle*. A composite bundle is a collection of any mobility and non-mobility goods within a creator's endowment.

## Utility Maximization Subject to a Mobility Constraint

Given some mobility creator with a mobility function:

$$M(L, K) = a L^\alpha K^\beta + b l^\gamma + c k^\delta$$

And a utility function:

$$U(M, A) = M^\phi A^{1-\phi}$$

Utility can be expressed a composite function:

$$U(M(L, K), A) = (a L^\alpha K^\beta + b l^\gamma + c k^\delta)^\phi A^{1-\phi}$$

A rational, utility maximizing creator will allocate his or her composite bundle such that:

$$\frac{\partial U}{\partial M} = \frac{\partial U}{\partial A} \quad \text{thus} \quad \frac{\partial U}{\partial M}\frac{\partial M}{\partial K} = \frac{\partial U}{\partial A} \quad \text{and} \quad \frac{\partial U}{\partial M}\frac{\partial M}{\partial L} = \frac{\partial U}{\partial A}$$

Which reduces to:

(6) $\quad \dfrac{\partial U}{\partial K} = \dfrac{\partial U}{\partial L} = \dfrac{\partial U}{\partial A}$

Equation 6 says that a creator will maximize his or her utility such that the marginal utility products of mobility (the marginal utility attained from hiring one additional unit of labor and capital toward attaining the next mobility good) are equal to the marginal utility of all other goods.

## Moving Along A Mobility Preference Curve vs Moving to a Larger Endowment

Suppose we have a creator who is contemplating purchasing an $8,000 electric wheelchair instead of continuing to use their existing $2,000 manual chair. Whether or not such a move is utility maximizing depends on the creator's mobility function and the value of the parameter $\phi$. If the creator finds that he or she is capable of getting around in the manual wheelchair without suffering any fatigue or repetitive strain, then the use of an electric chair would not be mobility maximizing. the intuition here is that the shoulders are a well-functioning piece of mobility creating machinery, albeit organic, that will be left underused. The electric wheelchair would serve to needlessly replicate a component of the creator's productive mobility capacity. We can refer to this as the *crowding out effect*, where organic bodily functions are crowded out by the over use of synthetic inputs. In this case, the $6,000 net expenditure would not generate any increases in mobility, which would put the creator at a lower level of utility because the non-mobility endowment shrank.

However, If the creator's mobility function was such that the electric chair reduced pain and inflammation caused by repetitive strain from pushing around all the time, the mobility endowment would grow. If the increase in utility were enough to surpass the $6,000 cost, the creator would move to a higher level of utility. If the change in utility were exactly equal to to the $6,000, the creator would be indifferent, which would generate a movement along the curve.

## III. Efficiency in Mobility Product Mix

**Attainment of the Pareto Set**

Let $\mathbb{M}$ be the set of all mobility functions of the form $M(L, K)$

Let $\mathbb{U}$ be the set of all mobility preference functions of the form $U(M, A)$

Let $\mathbb{A}$ be the set of all non-mobility production functions of the form $A(L, K)$

One would expect a Pareto dominating equilibrium where:

(7) $$\frac{\frac{\partial U}{\partial M}}{\frac{\partial U}{\partial A}} = \frac{\frac{\partial M}{\partial L}}{\frac{\partial A}{\partial L}} = \frac{\frac{\partial M}{\partial K}}{\frac{\partial A}{\partial K}} \quad \text{for all objects contained in } \mathbb{M} \cup \mathbb{U} \cup \mathbb{A}$$

Equation 7 describes an equilibrium called *efficiency in mobility product mix.* There is some unique allocation of mobility and non-mobility goods such that the ratio of marginal utilities for consumers are equal to the ratio of marginal products attained in producing the two goods. Such an equilibrium ensures that there is no re-allocation that would improve somebody's endowment without reducing the endowment of somebody else. The remainder of this section will explain how certain provincial disability support systems in Canada interfere with the attainment of such an equilibrium.

**inefficiencies Caused by Restrictions on Capital Allocations**

The provincial governments of Alberta and Ontario both have programs that use public funds to purchase mobility aides for creators. Alberta uses approved product lists.[5] Approved product lists contain mobility aides specific to make and model that the agency will purchase. If the specific product desired by a creator is not on the list, it will not be purchased. Ontario does not make purchasing decisions based on make and model, but they do restrict what they will purchase based on the type of mobility aid (i.e. they do not fund commode chairs). The problem is that these policies tend to prohibit the use of a great many mobility aides that would generate Pareto improving moves for creators.

---

[5] Alberta Aids to Daily Living, General Policy & Procedures Manual (February 2020)

One example of this can be found in the exclusion of turner-transfer-aide devices from the Alberta Aides to Daily Living (AADL) program. Turner-transfer-aides work well for creators who posses enough organic mobility to stand up autonomously, but cannot pivot their body while standing to enable movement from surface to surface. The only other option available is a ceiling track lift, which lifts the entire body from surface yo surface. Use of such a device would not be mobility maximizing because it would crowd out the use of the creator's legs. A lift might cost $2,000. A turner-transfer-aide costs $1,000. Just like in the preceding example, an additional net expenditure on mobility capital ($1,000 more to use a lift over a turner) does not cause the mobility endowment to grow, resulting in a reduced endowment.

One reason why Alberta uses approved product lists instead of restricting purchases by type is because AADL recycles all of the equipment they purchase. Difficulties arise with turner-transfer-aides because relatively few people use them. Most people who possess the organic mobility to stand autonomously can also shift their weight well enough using a ceiling mounted poll. This creates difficulties in finding another user for the device in the event that the user it was purchased for no longer needs it. The Ontario Assistive Devices Program (OADP) does not recycle mobility aides; users can apply for new equipment every so many years, and keep the old stuff. This greatly reduces the demand for used equipment, resulting a surplus of mobility capital. Markets are good at resolving such inefficiencies. If mobility creators were simply given a monetary endowment to spend as they saw fit, they would simply buy whatever mobility aide that enriched their endowment most.

**Inefficiencies Caused by Restrictions on Labor Allocations**

Another source of inefficiency prevails in home care support programs that allow mobility creators to hire caregivers. Sadly, the funds provided don't offer enough flexibility for households to make productivity maximizing allocations. Many households with mobility creators face a demand for *micro-allocations*. A micro-allocation is a very tiny use of labor employed to attain a mobility good that would otherwise be done autonomously. Suppose a creator is watching TV and drops the remote. They might ask a member of their household to pick it up. In the absence of someone to do so, a creator is often left with a reduced endowment because the alternatives tend to be energy consuming. A reaching devise, for example, might weigh up to five pounds, at it often takes two or three attempts at lifting it before success. This would shrink most creator's endowments because daily stamina is often very finite

Micro-allocations tend to be disruptive to the completion of other domestic tasks, like housecleaning and laundry, because of the interruptions they impose. If households were given the freedom to outsource these tasks instead, the result would be a more structured division of labour that would boost productivity. Surely it would be more productive for one labourer to focus on mobility creation, and another to the day-to-day chores of the household, instead of putting mobility creation solely in the hands of the fixed schedules of third parties, which leaves family members having to play both roles when the paid caregiver leaves.

One commonly cited reason for the reluctance of governments to allow monetary compensation for family caregivers is that it exacerbates the vulnerability of mobility creators within families; family members would have an incentive to spend less of the money on supports and more on luxury goods, for example.

# IV. Inter-temporal Mobility Creation and the Pareto Set

**Inter-temporal Mobility Functions**

Suppose a mobility creator were to enrol in a physical rehabilitation program that improved his or her organic mobility. The pre-rehab period will be denoted as $t_0$ and the post rehab period as $t_1$.

In period $t_1$ the creator may notice significant changes in things like muscle pain and fatigue, by virtue of having better bodily function. The creator might find that he or she can tolerate longer intervals of sitting upright, before having to take a break and lay down. This means that the creator now has access to more mobility goods. There is more time to be spent upright and mobile. Additionally, the creator may be strong enough to transfer into their commode using a wall mounted pole instead of a lift. The net result is a simultaneous increase in the creator's the mobility and non-mobility endowments.

Such changes can be shown in the creator's mobility function by treating the coefficients and organic parameters as functions of *t*. In this case, everything is a positive function of *t*; the increase in the number of mobility goods results in larger coefficients, and the fact that less is being devoted to capital expenditure for the same bundle means that organic parameters have increased.

(8) $\quad M(L, K, t) = a(t)L^{\alpha(t)}K^{\beta(t)} + b(t)l^{\gamma(t)} + c(t)k^{\delta(t)}$

Equation 8 is an example of an *inter-temporal mobility function*. Wherever there is an increase in organic mobility, the endowment coefficients and organic parameters are positive functions of time. Wherever there is a decrease in organic mobility, the organic parameters and endowment coefficients are negative functions of time.

**Type 1 and Type 2 Mobility Functions**

Mobility functions of the form expressed in Equation 6 can be classified into two types, based on the function's derivative with respect to *t*.

$M(L, K, t) = a(t)L^{\alpha(t)}K^{\beta(t)} + b(t)l^{\gamma(t)} + c(t)k^{\delta(t)}$

If $\dfrac{\partial M}{\partial t} \geq 0$, *M* is a Type 1 mobility function

If $\dfrac{\partial M}{\partial t} < 0$, *M* is a Type 2 mobility function

Where $\dfrac{\partial M}{\partial t}$ is the *rehabilitation rate,* or r-rate.

The absolute value of the rehabilitation rate tells us about the severity of the mobility impairment over time. A mobility creator with Multiple Sclerosis (MS) would have an r-rate that is farther away from zero than a mobility creator with Amyotrophic Lateral Sclerosis (ALS) because ALS is far more aggressive in its progression over time. Somebody with a spinal cord injury would have an r-rate close to zero because their organic mobility does not change over time. Somebody with an anoxic brain injury (Cerebral Palsy or Stroke) would have a positive r-rate because they can sometimes develop new motor-neural pathways that modestly enrich their organic mobility in the long-run. Type 1 mobility functions describe impairments that stay the same or improve over time, while Type 2 ones describe impairments that worsen over time.

**Inter-temporal Mobility Preference Functions**

Recall from Section II that a mobility preference curve was defined as

$$U(M, A) = M^\phi A^{1-\phi}$$

And that a total endowment was defined as

The derivation of the inter-temporal mobility function has shown that $E_M$ is a function of time, which means that $E_A$ is too, because of the endowment identity from Equation 5

$$E_T = E_M + E_A$$

The inter-temporal mobility function thus becomes:

(8) $\quad U(M, A, t) = M(t)^\phi A(t)^{1-\phi}$

The inter-temporal unity maximization criterion becomes:

(9) $\quad \dfrac{\partial U}{\partial M}\dfrac{\partial M}{\partial t} = \dfrac{\partial U}{\partial A}\dfrac{\partial A}{\partial t} \qquad PV(\dfrac{\partial U}{\partial M}\dfrac{\partial M}{\partial t}) = PV(\dfrac{\partial U}{\partial A}\dfrac{\partial A}{\partial t})$

A mobility creator will maximize inter-temporal utility by choosing an allocation such that the marginal utilities of mobility and all other goods with respect to time are the same. if time preferences are taken into account, the present value of the marginal utilities being the same would maximize utility. Some creators, for example, might have stronger preferences for current endowments when it comes to non-mobility goods, compared to those of mobility goods.

**An Example**

Suppose a Type 1 creator, who was a full-time electric wheelchair user, were to enter a rehabilitation program. After a few months of therapy, the creator might find that he or she has better hand function; they can put dishes away faster, and put on their coat on with less overall strain and fatigue. As wonderful as such improvements are, the creator would find that the electric wheelchair is problematic because it is so high off the ground that they cannot reach the floor to pick things up. One solution would be to buy a customized manual wheelchair to use indoors and for short distances outdoors.

The manual chair would yield many advantages beyond retrieving dropped items. The creator would have an easier time cooking because the lower chair would enable better stove access. The creator could get in and out of bed autonomously because it is much easier to move a manual wheelchair into a safe position, while sitting on the edge of a bed, compared to an electric; there is no joystick that is difficult to reach that must be used to move it. They can simply push the chair to wherever it is needed. The creator would weigh the present value of the additional future utility brought on by the chair against the cost of purchasing it today.

Upon trialing the new chair, the creator may form an expectation that they could go to bed autonomously if they were to continue rehab for a while in the new chair. This expectation is based on the fact that transferring in and out of bed can be done much more safely in the new chair. Once this is mastered, all one has to do is undress themselves to displace a care-giver. Given that the rehab has already generated improvements in putting on one's coat, it is perhaps reasonable to think that further gains in the area of undressing ought to happen. Suppose the wheelchair costs $6,000 and one year of rehab costs $3,000. After one year, the creator, having mastered the going to bed and undressing autonomously, does not need the additional one hour of caregiver time every night - at a cost of $25 per hour. This forgone expenditure generates $9,000 in annualized savings.

In this example, the creator spent $9,000 on the new wheelchair and one year of rehab in period $t_0$ and was able to re-capture the $9,000 in forgone mobility expenditures in period $t_1$, and all other subsequent periods. The creator also gets to enjoy an enriched mobility endowment where they can cook things easier and retrieve dropped items.

Such inter-temporal mobility improvements are difficult to attain because endowments aren't flexible enough to allow for changes in spending behaviour over time. Because caregivers and mobility aides are funded by different agencies with their own budgets, it would be impossible to spend less money on caregivers and keep that money to re-coup the costs of a new wheelchair. Furthermore, if the creator lived in Alberta, AADL would take the electric wheelchair away upon the arrival of the manual chair, because they only provide funding for one day-use chair. The loss of the electric wheelchair, for many creators, would result in so much fatigue and repetitive strain from having to push around everywhere that it would likely lead to a long-run reduction in total endowment.

If creators were simply given an annual monetary endowment to spend on mobility and all other goods, where the size of the endowment were contingent on the severity of the impairment, it would be much easier for them to save their money in one period to pay for more mobility expenditures in future periods. Another solution would be to create a financial instrument that would allow creators to borrow against the value of future savings so that they can spend more money re-tooling their mobility functions today. Under the current system, where reduced caregiver expenditures do not go back to the creators who innovate, there is a greatly reduced incentive for creators to explore new techniques of creating mobility.

**The Independence Premium**

Suppose, in our preceding example, that the creator saw no mobility endowment growth at all resulting from the new chair and the rehab. That is, he or she was not any better at cooking or retrieving things than before, and the only real benefit was the ability to go to bed autonomously, without the use of a caregiver. It would still be expected that the creator would get the new chair because of the future mobility expenditure savings induced by reduced caregiver use.

Suppose now that going to bed autonomously induced some fatigue for the creator. Suppose that the creator saw a reduction in mobility endowment that caused a loss of utility that was exactly equal to the present value of all the future income streams captured by the new methods. Would the creator still be enticed to buy the new chair? The answer would depend on the size of the creator's *independence premium*. An independence premium is a cost that a creator is willing to absorb, often associated with altering one's mobility function, when there is no endowment growth to offset it. For example, in the latter scenario where the reduction in mobility is exactly equal to the monetary value of the reduced expenditures, the creator would have no independence premium if he or she were indifferent between the new autonomous method or the old. However, if the creator expressed preferences toward buying the chair and doing the rehab, this would reflect an independence premium. The amount of time and effort that goes into the re-tooling of the production function is the premium that is being paid for the privilege of doing things on one's own

**Comparative Statics with the Inter-temporal Mobility Model**

Suppose a provincial government were to implement a new publicly funded physiotherapy program across a population of Type 1 mobility creators. Like before, we will denote the pre-therapy period $t_0$ and the post-therapy period $t_1$.

Let $\mathbb{M}_0$ be the set of all Type 1 mobility functions in period $t_0$ of the form:

$$M(L, K, t) = a(t_0)L^{\alpha(t_0)}K^{\beta(t_0)} + b(t_0)l^{\gamma(t_0)} + c(t_0)k^{\delta(t_0)}$$

We would expect an equilibrium $E_0$ of the form:

$$\frac{\frac{\partial U}{\partial M}}{\frac{\partial U}{\partial A}} = \frac{\frac{\partial M}{\partial L_0}}{\frac{\partial A}{\partial L}} = \frac{\frac{\partial M}{\partial K_0}}{\frac{\partial A}{\partial K}}$$

Let $\mathbb{M}_1$ be the set of all Type 1 mobility functions in period $t_1$ of the form:

$$M(L, K, t) = a(t_1)L^{\alpha(t_1)}K^{\beta(t_1)} + b(t_1)l^{\gamma(t_1)} + c(t_1)k^{\delta(t_1)}$$

We would expect an equilibrium $E_1$ of the form:

$$\frac{\frac{\partial U}{\partial M}}{\frac{\partial U}{\partial A}} = \frac{\frac{\partial M}{\partial L_1}}{\frac{\partial A}{\partial L}} = \frac{\frac{\partial M}{\partial K_1}}{\frac{\partial A}{\partial K}}$$

Does $E_1$ Pareto dominate $E_0$? The answer is yes. In the previous two examples, it was shown that the creator's endowments grew because the monetary value of forgone mobility expenditures and more mobility goods was greater than or equal to the cost of acquiring more organic mobility. Therefore, if a population of Type 1 creators are given access to rehabilitation, each individual will see an increase in his or her endowment, either in the form of reduced mobility expenditures, greater mobility goods, or both. The creator will continue spending on rehabilitation until such time that the additional utility generated by the mobility gains are less than the forgone utility of the enrolment fees had they been spent elsewhere.

The same maximization rule is also 2 across Type 2 populations. A Type 2 mobility creator might decide, under the care of a physician, to take a prescribed drug that helps to slow the pace of degeneration. The creator would continue taking the drug until such time that the loss of utility associated with the mobility endowment is equal to the forgone utility in all other goods induced by taking the drug.

Type 1 and Type 2 mobility creators will therefore simultaneously maximize their total endowments by adopting mobility expenditures that will generate future endowment growth, resulting in long-run Pareto dominant equilibria.

(10)    As $t \to \infty$    $E \to E*$    $\dfrac{\frac{\partial U}{\partial M}}{\frac{\partial U}{\partial A}} = \dfrac{\frac{\partial M}{\partial L*}}{\frac{\partial A}{\partial L}} = \dfrac{\frac{\partial M}{\partial K*}}{\frac{\partial A}{\partial K}}$    where $E*$ Pareto dominates $E$.

The intuition behind Equation 10 is simple. An economy where no individuals had any mobility impairments would Pareto dominate one where some people did. In the latter scenario, everyone's total endowment would be larger because they would not have to devote any purchasing power toward mobility creation; the body's organic structures would naturally generate all the mobility needed, which is tantamount to having an infinitely large mobility endowment. Therefore, any equilibria where there is more functional organic mobility to displace the synthetic Is Pareto dominant.

**Policy Induced Barriers to Efficiency**

The Canadian social safety net is somewhat better equipped to care for Type 2 creators compared to Type 1's because Pharmaceuticals are more ubiquitously available than rehabilitation programs. While rehabilitation is available in paediatric populations, and when an adult becomes newly disabled, they tend not to be as widely available for adult Type 1 creators with a stable prognosis. The result is an outcome where the Type 2 creators attain an equilibria that is far closer to the Pareto dominating long-run one than than Type 1 populations.

One explanation for this is what disability commentator Jim Mansell calls "the rise of market forces" fuelling such policy decisions.[6] Mansell argues that the rise of taxpayer centric decision making on healthcare and social programs has led to a doctrine where mobility creators are given just the bare minimum level of support required for their immediate survival. Because the pharmaceutical products available to Type 2 creators is ultimately a life extending therapy, they are deemed essential. Physiotherapy interventions, on the other hand, are not. The real tragedy here is the long-run loss of efficiency prevailing in adult Type 1 populations. There are a great many such individuals that could be capturing substantial savings in forgone mobility expenditures tomorrow if they had access to a more diverse pool of interventions today.

## V. Accessibility and Incentive Compatibility

This section will add one final layer to the model thus far. It will discuss what happens to the behaviour of mobility creators, and consequently the Pareto Set, when society starts making things accessible in the community.

**Demand for Accessibility**

Suppose we have a creator with a mobility preference function: $U(M, A) = M^\phi A^{1-\phi}$

In a society where nothing were made accessible (i.e. not cut curbs to allow wheelchairs onto sidewalks, no elevators to use in lieu of stairs, no accessible washrooms) the creator's utility would be constrained a great deal. Once their caregiver leaves after the shower dress routine is done in the morning, the creator would not be able to go out and do anything; they could not drive their wheelchair to the drugstore, or take a bus to the mall. In this case their would be a reduced utility associated with all mobility and non-mobility goods. Having more money is meaningless if one cannot go out to spend it. Having more hours in a day sitting upright doesn't mean as much if you must spend them inside your house.

---

[6] Jim Mansell, "Deinstitutionalization and Community Living: Progress, Problems and Priorities," *Journal of Intellectual and Developmental Disability* vol. 31, no. 2 (2006): 65-76.

An outcome where things are accessible Pareto dominates one where things are not; all mobility creators are now attaining a higher level of utility by going out into their communities.

We will cal the initial Pareto set where noting is accessible $\mathbb{U}$, and the dominant one with accessibility $\mathbb{U}^*$. Unfortunately, $\mathbb{U}^*$ will only Pareto dominate $\mathbb{U}$ when there are no costs associated will re-tooling all of society's infrastructure, transportation and buildings to make them accessible. $\mathbb{U}^*$ is often difficult to attain for this reason.

**Kaldor-Hicks Efficiency and the Problem of Cost**

Markets are not very good at ensuring accessibility because creators cannot transact with business owners very efficiently. It would be very difficult for a single business owner to ask 10,000 mobility creators to each give he or she 1 dollar to pay for an elevator to be installed. Such transaction costs would be very high to do this for every building and the free-rider problem would limit the frequency of success with such initiatives.

Under such conditions where transactions cannot generate socially efficient outcomes, the next best thing is to strive for *Kaldor-Hicks efficiency*. Kaldor-Hcks efficiency says that welfare maximizing outcomes will be attained when the total social benefit of something exceeds the total cost. To determine if making communities accessible is welfare maximizing, one must determine if the total social benefit (the increase in utility for all the creators going out) is outweighed by the cost of installing elevators, modifying sidewalks and so on.

**Modelling the Costs and Benefits**

We will adopt a new parameter $\rho$, called the *Relative Accessibility Multiplier*. This multiplier, defined between $0$ and $1$, is an approximate proportion of how many things in a given community are accessible relative to the total. Therefore, by integrating a mobility preference function with respect to $\rho$ we get:

$$U = M^\phi A^{1-\phi} \qquad \int_0^1 U d\rho = \rho M^\phi A^{1-\phi}$$

(11) $\qquad U(M, A) = \rho M^\phi A^{1-\phi} \qquad$ where $0 > \rho \geq 1$

Equation 12 says that the utility attained by any given creator is a scalar multiple of the relative level of community accessibility.

Let $\mathbb{U}$ be the set of all functions $U = \rho M^\phi A^{1-\phi}$ where $\rho = 0.1$

Let $\mathbb{U}^*$ be the set of all functions $U = \rho M^\phi A^{1-\phi}$ where $\rho = 0.99$

The total social benefit of transforming the infrastructure in a community from minimally accessible $\mathbb{U}_{0.1}$ to highly accessible $\mathbb{U}_{0.99}$ is thus:

$$\int_{0.1}^{0.99} \mathbb{U} d\rho = 0.99 M^\phi A^{1-\phi} - 0.1 M^\phi A^{1-\phi}$$

Equation 13 generalizes this result, and then discounts the resulting utility change to account for the fact that the utility gains generated will continue into future periods.

$$(12) \quad \int_a^b \mathbb{U} d\rho = TC \big|_a^b \qquad\qquad PV(\rho M^\phi A^{1-\phi})\big|_a^b = TC\big|_a^b$$

Equation 14 shows the marginal cost relationship corresponding to the total cost one above. The idea is that policymakers will adopt accessibility expenditures such that the marginal cost of each additional project is equal to the present value of the utility gains it generates.

$$(13) \quad \Delta U = MC \qquad\qquad PV(\Delta U) = MC$$

Equation 15 shows that the marginal cost of an accessibility expenditure can approach zero, as can its corresponding utility. One example of this would be the installation of a small wooden ramp used to scale an obstacle of two inches at a local storefront. In relative terms, this could be interpreted as an infinitesimally small expenditure.

Let $b - a = h$

$$(14) \quad \lim_{h \to 0} \int_a^{a+h} \mathbb{U} d\rho = MC \qquad\qquad PV(\lim_{h \to 0} \int_a^{a+h} \mathbb{U} d\rho) = MC$$

A municipality might, for example, determine that it is Kaldor-Hicks improving to make the city's sidewalks and subway cars accessible because their relatively high use guarantees that enough creators will use them and hence the cost will be offset. They would not spend public money on making bars and restaurants accessible because the marginal cost of renovating each building would exceed the relatively small gains in utility that each renovation would generate. The decision of whether or not to make such renovations falls on the individual proprietor. Many such establishments tend not to be accessible because the costs associated with becoming accessible are not offset by very much increase in overall profit that result from accessibility.

The resulting system of *fragmented accessibility* is a source of great uncertainty for creators. Buildings that are constructed after the municipality's building codes were updated are accessible, but ones that were constructed under the outdated codes that did not prioritize accessibility are not. The alternative is to build communities that adopt *homogenous accessibility*, which more public money is spent to make everything accessible, as has been done in the United States,[7]

**Inefficiencies Induced by Fragmentation**

Suppose a creator were trying to decide if he or she should spend money on a customized wheelchair that would make it easier to navigate curbs and sidewalks, or get under low tables at restaurants. If there were a great deal of uncertainty with respect to how many places were accessible that the wheelchair could be used at, and how much one would even enjoy frequenting these places, would the purchase still be worthwhile?

---

[7] Americans With Disabilities Act of 1990, Public Law 101-336, 108th Congress, 2nd session (July 26, 1990).

The information costs that such things impose on creators would likely have some bearing on their decisions of input choice. The status quo bias[8], when applied in this context, predicts that creators would be more likely to keep their current input choice the same instead of purchasing new equipment. Recall from Equation 6 that creators maximize their short-term utility such that $\frac{\partial U}{\partial K} = \frac{\partial U}{\partial L} = \frac{\partial U}{\partial A}$.

In the presence of fragmentation, the additional information costs imposed on creators could be such that they never realize their maximum level of utility. As an intrinsic consequence of this behaviour, one could expect that a population of creators may never attain the Pareto set induced by the equilibrium $E \; \frac{\frac{\partial U}{\partial M}}{\frac{\partial U}{\partial A}} = \frac{\frac{\partial M}{\partial L}}{\frac{\partial A}{\partial L}} = \frac{\frac{\partial M}{\partial K}}{\frac{\partial A}{\partial K}}$.

Analogously, recall from Equation 9, that in the long-run creators maximize utility such that

$$\frac{\partial U}{\partial M} \frac{\partial M}{\partial t} = \frac{\partial U}{\partial A} \frac{\partial A}{\partial t}.$$

A creator deciding whether or not to enrol in a physiotherapy program with the goal of displacing the need for a caregiver in the evening, for example, would struggle with the same informational burdens. If it were difficult to determine which places were accessible, and hence where one could spend one's evenings, would the time, money and effort required to displace the caregiver and come home whenever one wanted really be worth it?

Just like in the short-run, such systemic uncertainty across a population would thus make the Pareto set induced by the equilibrium $E * \frac{\frac{\partial U}{\partial M}}{\frac{\partial U}{\partial A}} = \frac{\frac{\partial M}{\partial L *}}{\frac{\partial A}{\partial L}} = \frac{\frac{\partial M}{\partial K *}}{\frac{\partial A}{\partial K}}$ difficult to attain.

This loss of surplus induced by fragmentation will be called *residual inefficiency* denoted as $R$. The new welfare maximization criterion, under fragmentation, thus becomes:

(15) $\quad \int_a^b \mathbb{U} d\rho = TC \rvert_a^b + R$

## VI. Labor Supply and Mobility Creation

Thus far, this paper has assumed zero labor force participation by mobility creators. This section will relax that assumption, and show how mobility creation affects labor supply decisions for creators participating in the labor force.

---

[8] William Samuelson and Richard Zeckhauser, "Status Quo Bias in Decision Making" *Journal of Risk Uncertainty,* vol. 1 (1988) 7-59.

**The Consumption, Leisure and Mobility Trade-Off**

A mobility creator must decide how much time he or she is willing to spend on mobility creation in the short-term, perhaps a 16 hour day, for example. Of those 16 hours, some will be devoted to mobility creation $M$, and the rest will be devoted to non-mobility goods, or all other goods $A$. Mobility is a function of consumption $c$, and time $t$. All other goods is a function of consumption $c$, and leisure $l$. One can now construct a more sophisticated utility function of the form:

(16) $$U(M, A) = M(c(t), t), A(c(t), l(t))$$

A creator will maximize utility by selecting some value of $t$ such that

# VII. Conclusions

This paper has shown that there are allocations that mobility creators could make in the short-run that would be Pareto improving, if they were given fewer constraints with their input choices. It has also shown that inter-temporal mobility creation results in mobility functions that are highly fluid, with parameters that can change in the long-run.

Thus, welfare maximizing policies are ones that seek to relax constraints on input choice so that the Pareto set can be attained in the short-run. In the long-run, creators must be given more freedom so that they can alter their mobility expenditures today in order to grow their endowments in the future.

These outcomes will be difficult to accomplish with fragmented accessibility and a patchwork of autonomously funded and managed services (i.e. one agency for caregiver provision, another for mobility aide purchases) and so on. Mobility creators need a fungible endowment that they can spend as they see fit, where the goal will be to minimize uncertainty and encourage creators to attain the highest level of mobility they can, subject to the severity of their impairment and the level of monetary support that society is comfortable giving them. Such a model can only exist if there is an abandonment of the doctrine that supports for the mobility impaired must only enrich mobility endowments. It will also require an abandonment of the idea that everything need not be accessible in society.

What is mobility? Mobility is the universal facilitator, the universal compliment, that enables the consumption of every and all good or service, either in the form of functioning bodily structures, or synthetic inputs. A good disability support system, therefore, is one which ensures that as few of society's resources are devoted to facilitating consumption so that the rest of all that remains can be consumed to the fullest.The only true way to accomplish this is to let self-interested mobility creators try things with the resources of their own endowments and enjoy the fruits of their experiments.

## VIII.Works Cited